\documentstyle[preprint,aps,epsfig] {revtex}
\tighten
\begin{document}
\draft
\title{Local and non-local  equivalent potentials
for p-${}^{12}$C scattering}
\author{A. Lovell and K. Amos}
\address{ School of Physics, University of Melbourne,\\
 Victoria 3052,
Australia.}
\maketitle
\begin{abstract}
A Newton-Sabatier fixed energy inversion scheme 
has been 
used to equate
inherently non-local p-${}^{12}$C
potentials at a variety of energies to pion threshold,
 with
exactly  phase equivalent local ones.
Those energy dependent local potentials
then have been recast in the form of non-local
Frahn-Lemmer interactions.
\end{abstract}

\section { Introduction}

Over the years, local nuclear optical potentials have
been used predominantly as the interaction potentials
between colliding nuclei.
Frequently, those local interactions were specified
phenomenologically as 
Woods-Saxon (WS) optical potentials;  
the parameter values of which
were determined by variation
to find best fits to
nuclear elastic scattering data.
The geometries of those WS forms
often were  taken
commensurate with known attributes of nuclear densities
and/or to be
consistent with the character of Hartree fields for the nuclei.
Most importantly in many cases, quality
fits to scattering data were obtained using these local potential forms,
with parameter
values that varied smoothly with energy and mass.
In reality however, as the optical potentials can be specified in many nucleon
scattering theory as a folding of pairwise nucleon-nucleon $NN$ 
interactions with nuclear structure,
they must be non-local.
That is assured by the
Pauli principle which implies that the scattering theory
involves nucleon exchange amplitudes.

For nucleon-nucleus ($NA$) elastic scattering,
it is now possible to define 
non-local optical potentials in coordinate space
and to use them without recourse
to localization techniques~\cite{Review}.
Solutions of the attendant integro-differential
forms of the Schr\"odinger equations with those
non-local $NA$ potentials have given good to excellent
fits to cross-section and analyzing power data for a wide
range of target masses and for energies 40 to 800 
MeV~\cite{Review,set,Deb}.
Those non-local $NA$ potentials are formed by folding
effective, medium dependent, two-nucleon ($NN$) interactions 
with the one body density matrices (OBDME) of the target nucleus.
Very good results for light mass nuclei in particular
have been obtained when those OBDME were specified 
from shell model (or similar) calculations involving
very large and complete shell spaces.

Nevertheless, for a multitude of uses, such as with 
application of multi-step reaction theories and scattering 
into the continuum,
it is useful to define a local form of optical potential for the 
$pA$ scattering system.
Note also that
equivalent local potentials can be
used to account for the coupled channels effects
in scattering~\cite{Fi92},
the
resulting form related to that found by localization of 
exchange amplitudes.
There are many ways in which a local potential
can be specified and considered equivalent
to a non-local one. Some have been reviewed
recently~\cite{Review}.
Herein we consider a form that is to be phase equivalent.
The phase shifts provided by 
the $g$-folding (non-local) potentials used recently~\cite{Deb}
to find very good predictions of p-${}^{12}$C elastic scattering,
have been used 
as input to
finding solutions of fixed energy quantum inverse 
scattering theory. In particular
we have found the equivalent local potentials by using
a modified version of the Newton-Sabatier inversion scheme
that has been proposed by Lun {\it et al.}~\cite{Lun}. 
This method allows extraction of both central
and spin-orbit interactions from the input
phase shift values.
It is not the only fixed energy method that can do so.
Hooshyar~\cite{hooshyar} has used the Sabatier
interpolation formulae in a finite difference inversion method,
and Huber and Leeb~\cite{huber} have investigated an approach
to this problem based upon Darboux transformations.
Likewise an approximate scheme~\cite{4LeebHF}
has been used with some success~\cite{laa} to analyze
neutron-alpha particle scattering data in particular.
However, the scheme we adopt is most facile,
reducing the process
of inversion to finding solution of a system of linear-algebraic
equations.

Fixed energy inverse scattering schemes are
not the only ways to effect
the transition from non-local potentials to equivalent local ones
and {\it vice versa}.
Various other approximations and analytic
expressions exist~\cite{Review}.  However,
all such other methods rely upon the
range of the non-locality being 
small in
comparison to the size of most nuclei.  If such is also the case with
regard to the de Broglie wavelength of the projectile, a
semi-classical WKB approximation may be valid.  However, Peierls and
Vinh-Mau~\cite{Pe80} note that while the localization approximations
should be well met within the medium of a large nucleus, corrections
could be important in the nuclear surface region and for light mass
nuclei.  In any circumstance that the non-locality range exceeds the
characteristic length of the system, they note that the non-local
potential cannot be approximated by their algorithm of
localization.
 A desirable feature of using inverse scattering 
theory to define equivalent local potentials
is that none of these approximations are invoked.
But such must be borne in mind when 
the reverse process, of defining a utilitarian
non-local form from the energy variation of a set of 
local potentials, is considered.
In doing that reverse process, we chose 
a mapping used by Apagyi {\it et al.}~\cite{Apagyi}
for its  simplicity and
as the data sets 
we study exist
at disparate energies in the range 65 to 250 MeV.

In this paper the salient features of 
the modified 
Newton-Sabatier fixed energy inversion scheme are discussed first.
Then in Sec.~III, we present and discuss
the results of this
non-local to phase equivalent local potential scheme.
The  method 
to recast non-local 
forms from energy variation of
local potentials is then defined in Sec.~IV,
and 
results of mapping the local optical potentials
back to non-local interaction forms of Frahn-Lemmer type~\cite{Frahn}
are given then in Sec.~V.
Conclusions are drawn in Sec.~VI.

\section{The modified Newton-Sabatier fixed energy inversion scheme}

The modified Newton-Sabatier method~\cite{Lun}
permits extraction from a set of scattering phase shifts,
not only of a central local interaction
between two colliding quantum objects but also
a spin-orbit term.
The approach
relies on analogues of the Regge--Newton equations for
interactions involving the spin-orbit component of the potential, namely
the Sabatier transformation equations~\cite{Sabpaper}. 
In this method, the
non-linear Sabatier interpolation formulae
essentially are converted
into a finite set of linear-algebraic equations 
which are easily solved.

With most inverse scattering theories,
the assumed equation of motion is the homogeneous
Schr\"{o}dinger equation, the radial terms of which
we consider in the form
\begin{equation}
\left[\frac{d^2}{dr^2} + \frac{2\mu E}{\hbar^2} 
- \frac{\ell(\ell + 1)}{r^2} 
- \frac{2\mu}{\hbar^2} V(r)\right]
 R_{\ell, s, j}(r)=0\ .
\label{Eqn1}
\end{equation}
$V(r)$ is the hadronic interaction potential which
we assume is a sum of central and spin-orbit terms, viz.
\begin{equation}
V(r)= \left[ V_c(r) + V_{so}(r)\ 
{\mbox{\boldmath 
$\ell \cdot \sigma $}}
\right] ,
\end{equation}
where { \mbox{\boldmath $\sigma$}} = 
2{\mbox{\boldmath
$s$}}.
With charged particle scattering, $V_c(r)$
also includes the Coulomb potential.

Since the energy is fixed it is convenient to recast 
Eq.~(\ref{Eqn1}) into
a dimensionless form, using the following
notation
\begin{eqnarray}
\rho = kr, \ \ \  k^2 = {2\mu E \over \hbar^2}, \ \ \ U_c(\rho)={V_c(r)\over E}
\ \ \ {\rm and} \ \ \ U_{so}={V_{so}\over E}.
\end{eqnarray}
The dimensionless, decoupled, reduced radial 
Schr\"{o}dinger equations subsequently are 
\begin{equation}
\left[\frac{d^2}{d\rho^2} + 1 - U_c(\rho) +
s U_{so}(\rho) \mp 2s \lambda U_{so}(\rho) -
{\lambda^2 - \frac{1}{4}\over \rho^2}\right]
\chi^\pm_\lambda(\rho) = 0
\label{schr}
\end{equation}
where $\lambda (= \ell + \frac{1}{2})$
 is the angular momentum variable.

As two unknowns, $V_c(r)$ and $V_{so}(r)$, are sought, only
two of the ($2s+1$) possible equations in
Eq.~(\ref{schr}) are required.
It is convenient to choose the cases for which
$j = \ell \pm s$ associated with which are two sets of phase shifts 
denoted by $\delta^+_\ell$ and $\delta^-_\ell$ respectively. This method
can be used with particles of any spin.
For $NA$ scattering, the nucleon intrinsic spin of course is $\frac{1}{2}$,
but 
we must presume that the spin of the nucleus is not an influential
factor in scattering, i.e. we equate all such scattering to that
from a spin zero target so that the quantum number is
$j = \ell \pm \frac{1}{2}$. 
This is precisely the case though for p-${}^{12}$C
scattering that we investigate herein.
The superscripts
`$\pm$' now designate the relevant two values of $j$.

The Sabatier interpolation formulae
relate the regular solutions of Eq.~(\ref{schr}),
$\chi^\pm_\lambda (\rho )$, to the set $\{\psi_\lambda(r)\}$, which are
solutions of the Schr\"odinger equations
for a reference potential, $U_0(r)$, 
that are regular at the origin.
The link is
\begin{equation}
\chi^\pm_\lambda(\rho)=F^\pm(\rho)\psi_\lambda(\rho)+
\sum_{\mu\in \Omega} {2\mu\over \pi} W_{\lambda\mu}(\rho)
\left[b_\mu\chi^\pm_\mu(\rho)-a^\pm_\mu\chi^\mp_\mu(\rho)
\right]\
\label{inter}
\end{equation}
where
\begin{equation}
W_{\lambda\mu}(\rho)={\psi_\mu(\rho)\psi_\lambda^\prime(\rho)-
\psi_\lambda(\rho)\psi_\mu^\prime(\rho)\over \lambda^2-\mu^2}
\end{equation}
is the Wronskian. The
set of scalar functions, $F^\pm$, in Eq. (\ref{inter}) are defined by
\begin{eqnarray}
F^\pm(\rho) &=& \exp\left[\pm\int^\rho_0 t S U_s(t)dt\right]
\nonumber\\
&=& {2\over\pi\rho}a_0\chi_0(\rho)\psi_0(\rho)+
{2\over\pi\rho}\sum_{\mu \in S}\left[a^\pm_\mu
\chi^\mp_\mu(\rho)+b_\mu\chi^\pm_\mu(\rho)\right]\psi_\mu(\rho)\
\label{Fpm}
\end{eqnarray}
and obey the following property:
\begin{equation}
F^+(\rho) F^-(\rho) = 1.
\end{equation}

Solution of these equations rely on a complete
knowledge of all the phase shifts, $\delta^\pm_\lambda$,  where 
$\lambda  \equiv \ell + 0.5$
now includes not only the set of
corresponding 
to the physical set of integer $\ell$ but also all
half-integer values. As such, the summations span the set of angular
momenta $\Omega :\{{1\over 2},\ 1,\ {3\over 2},\ 2,\ {5\over 2},\ 3,\
\cdots\}$.
Naturally, analysis of scattering 
data can provide only the physical set of phase 
shifts, i.e.
those corresponding to
integer $\ell$. 
Interpolation of  the (physical) data set 
is then required to 
define the required set for inversion.
Of course if instead one wishes to map against a defined
non-local potential, then the phase shifts at the intervening
values for $\lambda$ can be evaluated.

Eqs. (\ref{inter}) and (\ref{Fpm}) contain a set of unknown coefficients
$a_\lambda^\pm$.
Eq. (\ref{Fpm})  also includes functions $\chi_0(\rho)$ and $\psi_0(\rho)$
which are solutions of the
relevant Schr\"{o}dinger equations for $\ell = -{1\over 2}$. In addition
there are the  weights $b_\lambda$ which are defined~\cite{CSbook}
by
\begin{equation}
b_\lambda =
{\gamma(\lambda+{1\over2} - i\eta)\gamma(\lambda+{1\over2} + i\eta)\over
\left[ \gamma(\lambda + {1\over2}) \right]^2}
\left\{
\begin{array}{ll}
\phantom{-}
\cosh(\pi\eta) & {\rm integer}\ \lambda\nonumber\\
- \sinh(\pi\eta) & {\rm half-integer}\ \lambda
\end{array}
\right.\ .
\label{b}
\end{equation}

The set of equations, Eqs.~(\ref{inter}) and (\ref{Fpm}), constitute
a matrix problem to specify the unknown coefficients,
 $a_\lambda^\pm$,
and the functions, $F^\pm$, when select large
radial values $\rho_i$
are chosen so that
\begin{eqnarray}
\psi_\lambda^\pm(\rho_i) &=& \sin
\left[ \rho_i - \frac{1}{2} \left( \lambda - \frac{\pi}{2}\right)
+ \sigma_\lambda - \eta \ln(2\rho_i)
\right]\nonumber\\
\chi_\lambda^\pm(\rho_i) &=& c_\lambda^\pm \sin
\left[ \rho_i - \frac{1}{2} \left( \lambda - \frac{\pi}{2}\right)
+ \delta_\lambda^{\pm} - \eta \ln(2\rho_i)
\right]\nonumber\\
F^\pm(\rho_i) &=& h^\pm;\ \ {\rm (constants)}\ ,
\end{eqnarray}
where $\sigma_\lambda$ are the Coulomb phase shifts
and $c_\lambda^\pm$ are additional unknown constants to be 
determined.
Taking two (or more) distinct radial values,
 ($\rho_i = \rho_1, \rho_2, \cdots \ge \rho_0$),
and with $\chi_\lambda^\pm(\rho) \equiv c_\lambda^\pm
T_\lambda^\pm(\rho)$, Eqs.~(\ref{inter}) and (\ref{Fpm})
can be recast as
\begin{equation}
\psi_\lambda(\rho_i)
= C^\pm_\lambda T^\pm_\lambda(\rho_i) +
\sum_{\mu\in S^\prime}
{2\mu\over\pi} W_{\lambda\mu}(\rho_i)
\left[T^\mp_\mu(\rho_i) A^\pm_\mu -
b_\mu T^\pm_\mu(\rho_i) C^\pm_\mu\right]\ .
\label{interfree}
\end{equation}
They form a set of linear equations in the unknown
coefficients which have been grouped as
\begin{equation}
A^\pm_\lambda 
= a^\pm_\lambda c^\mp_\lambda h^\mp\ \ \ \mbox{and}
\ \ \
C^\pm_\lambda = c^\pm_\lambda h^\mp\ .
\label{abb2}
\end{equation}
The set $S^\prime$ is limited to $\lambda_{max}$
so that the matrix is finite 
and we can find a solution 
by using singular value decomposition
(SVD); a useful approach since the matrix may tend
to be ill-conditioned.
By so doing 
we presume that from a 
characteristic radius, $\rho_0 = k r_0$, the interaction
is solely Coulombic (or zero for incident neutrons) and that 
for all $\lambda > \lambda_{max}(\rho_0), 
\ \ \delta_\lambda^\pm - \sigma_\lambda \to 0$.
With the coefficients $a_\lambda^\pm,\ {\rm and}\ c_\lambda^\pm$
defined ($h^\pm$ are specified in terms of them),
multiplication of Eq.~(\ref{inter}) by $F^+(\rho)$
and upon rearrangement, gives
a set of $(8*\lambda_{max} + 1)$ linear equations,
\begin{eqnarray}
\left[F^{+}(\rho)\right]^2 \psi_\lambda(\rho) =
F^{+}(\rho) {\chi}^{+}_\lambda(\rho) -
\sum_{\mu\in \Omega^\prime}
{2\mu\over \pi} W_{\lambda\mu}(\rho)\ \left[b_\mu
F^{+}(\rho) {\chi}^{+}_\mu(\rho) -
a^{+}_\mu F^{+}(\rho)
{\chi}^{-}_\mu(\rho)\right]\ \ \ &&\nonumber\\
\psi_\lambda(\rho) =
F^{+}(\rho) {\chi}^{-}_\lambda(\rho)-
\sum_{\mu\in \Omega^\prime}
{2\mu\over \pi} W_{\lambda\mu}(\rho)\ \left[b_\mu
F^{+}(\rho) {\chi}^{-}_\mu(\rho) -
a^{-}_\mu F^{+}(\rho){\chi}^{+}_\mu(\rho)
\right]&&\nonumber\\
\left[1-(F^+(\rho))^2\right]\psi_0(\rho)
=\sum_{\mu\in \Omega^\prime}{2\mu\over\pi}W_{0\mu}(\rho)
\left[(b_\mu+a^-_\mu) F^{+}(\rho) {\chi}^+_\mu(\rho)
- (b_\mu+a^+_\mu) F^{+}(\rho){\chi}^-_\mu(\rho)\right]\ ,
\end{eqnarray}
that can be solved for the $(8*\lambda_{max} + 1)$ values of
$F^{+}(\rho) \chi_\lambda^{\pm}(\rho)$ and
$F^{+}(\rho)$ at each value of $\rho \le \rho_0$ desired.
Then, as Chadan and Sabatier~\cite{CSbook}
have shown,
the inversion potentials
are obtained from identities
\begin{eqnarray}
U_{\rm so}(\rho) &=& \pm \frac{2}{\rho} 
\frac{d\ln(F^\pm(\rho))}{d\rho}
\ \ \ \mbox{with}
\ \ \ F^+(\rho) F^-(\rho) = 1\nonumber\\
U_{\rm c}(\rho) &=& U_0(\rho) + 
\frac{1}{2} U_{\rm so}(\rho)
- \frac{1}{\rho} 
\frac{d}{d\rho}
[G^+(\rho)F^-(\rho) + G^-(\rho)F^+(\rho)]
+ \frac{1}{4} [\rho U_{\rm so}(\rho)]^2\ ,
\label{Veqn}
\end{eqnarray}
where
\begin{equation}
G^{\pm}(\rho) =
{2\over\pi\rho}\sum_{\mu \in \Omega}\mu\left[ a^\pm_\mu
\chi^\mp_\mu(\rho)-b_\mu\chi^\pm_\mu(\rho)\right]\psi_\mu(\rho)\ .
\label{Geqn}
\end{equation}

Inherent with inversion potentials from all Newton-Sabatier methods
is a pole at the origin. That pole arises from the s-wave
contributions in the summations~\cite{MEpaper}
but it usually influences only small radii
properties of the results.  In most cases studied,
the effects of the pole are not evident beyond 0.5 fm
typically.

The Coulomb field poses a problem with inverse scattering
theory applications.  The scattering phase shifts 
with a Coulomb potential incorporated
increase with $\ell$.
However, a transformation of phase shifts allows that
problem to be allayed~\cite{May}.
At and beyond a radius $r_0$, the nuclear component
in the total potential can be ignored.
So at $r_0$ the Schr\"odinger potential is given by
 $V_c(r_0)={2\eta E/\rho_0}$.
The method is to take that value for 
a new potential at all larger radii,
 and hence
that new potential is
\begin{equation}
\tilde{V}(r) =
\left\{
\begin{array}{ll}
 V(r) &  r < r_0=\rho_0/k \\ \nonumber
 V_c(r_0) & r \ge r_0\\
\end{array}
\right.\ .
\label{modNS}
\end{equation}
For this potential, 
the internal solutions are then to be matched to Bessel 
functions (zero reference potential solutions).
Subtracting the long ranged constant potential $V_c(r_0)$ 
from the overall one, 
\begin{equation}
\tilde{V}(r)-V_c(r_0)=
\left\{
\begin{array}{ll}
 V_c(r) - V_c(r_0)  & r < r_0\\ \nonumber
 0                  & r > r_0\\
\end{array}
\right.\ ,
\end{equation}
gives a new potential which
when used in
the Schr\"{o}dinger equation gives rise to
a new but essentially equivalent set of phase shifts.
By matching the logarithmic derivatives 
at $r_0$ with the external, zero potential
solutions, one finds the relation
\begin{equation}
{d\over dr}\log[\cos\tilde{\delta}_\lambda H_\lambda(\beta\rho_0)+\sin
\tilde{\delta}_\lambda I_\lambda(\beta\rho_0)] =
{d\over dr}\log[\cos\delta_\lambda \psi_\lambda(\beta\rho_0)+\sin\delta_\lambda
\zeta_\lambda(\beta\rho_0)]\ ,\label{mayeq}
\end{equation}
\noindent where
$H_\lambda$ and $I_\lambda$ are the regular and irregular zero potential
solutions respectively
and $\beta=\sqrt{1-V(\rho_0)/E_{cm}}\ \ (= \sqrt{1-2\eta/\rho_0}$
for a Coulomb potential$)$.
The new phase shifts 
are then given by
\begin{equation}
\tilde{\delta}_\lambda= -\arctan\left({H^\prime_\lambda(\beta\rho_0)-H_\lambda
(\beta\rho_0)D_\lambda\over I^\prime_\lambda(\beta\rho_0)-I_\lambda(\beta\rho_0)
D_\lambda} \right)\
\end{equation}
where
\begin{equation}
D_\lambda(\rho_0)=i\left[{\cos\delta_\lambda
\psi^\prime_\lambda(\rho_0)+\sin\delta_\lambda\zeta^\prime_\lambda(\rho_0)\over
\cos\delta_\lambda\psi_\lambda(\rho_0)+\sin\delta_\lambda\zeta_\lambda(\rho_0)}\beta \right] .
\end{equation}
Thus inversion is made of these new phase shifts with a zero reference
potential to obtain
$\tilde{U}(\rho)$ from which and on accounting for 
the energy shift, the actual potential for $r < r_0$ is 
\begin{equation}
V(r)
=E \left[ \beta^2 \tilde{U}(r) + 1 -\beta^2
\right],\ \ \ \ \ \ \ \ r={\rho\over k\beta}
\end{equation}


\section{Results and discussion}

Using medium modified effective $NN$ interactions between
a projectile proton and each and every bound
nucleon
in ${}^{12}$C, microscopic non-local
complex and spin dependent optical potentials
have been generated for a range of
energies, 40 to 800 MeV in fact~\cite{Deb}.
The resulting cross sections and analyzing powers
so predicted are in good agreement 
with the observed data.
Those calculations were made 
using the program
DWBA98~\cite{Raynal}.
Therein, at each (fixed) energy,
the appropriate effective $NN$ interaction is folded with
the OBDME of the target.
 For ${}^{12}$C those OBDME were generated from complete
$(0+2)\hbar\omega$ space shell model wave
functions~\cite{set}.
That program also solves the 
integro-differential form of Schr\"odinger equations,
and thereby
we obtained sets of phase shifts of (non-local)
optical
potentials to
use as input to the inversion
procedure.

\subsection{The equivalent local potentials from inversion}

\begin{figure}
\centering\epsfig{figure=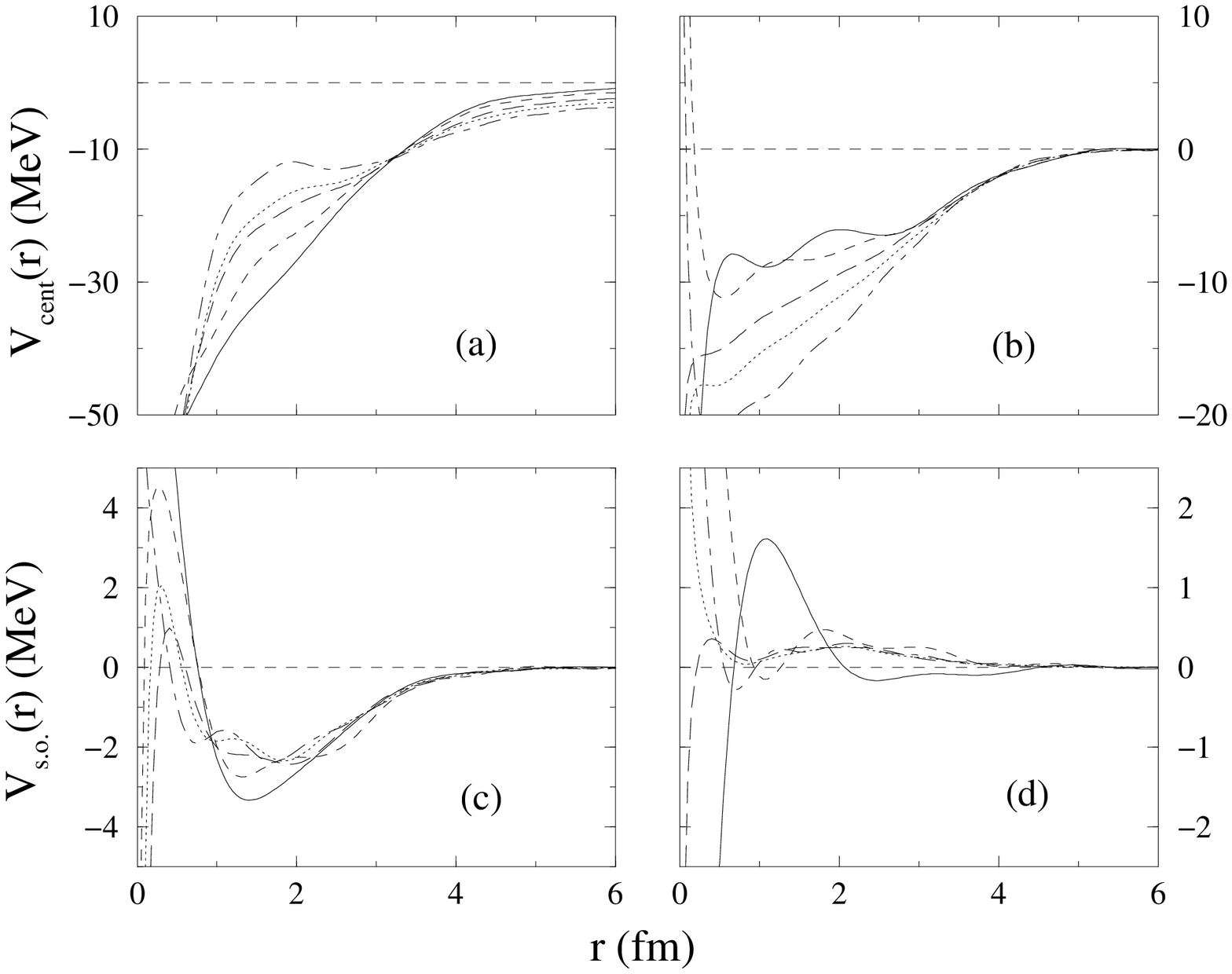,width=\linewidth,clip=}
\caption[]{
Inversion potentials from p-${}^{12}$C scattering analyses
portraying (a) the central real, (b) the central imaginary,
(c) the spin-orbit real, and (d) the spin-orbit imaginary components.
The separate energy results 
are identified in the text.
}
\label{pots}
\end{figure}

The inversion potentials obtained using the $g$-folding model phase shift sets
specified for proton scattering from ${}^{12}$C with
energies 
65, 100, 160, 200, and  250 MeV 
are displayed in Fig.~\ref{pots} by 
the solid, dashed, long dashed, dotted and dot-dashed curves
respectively.
Results have been found for 135 MeV protons
as well and 
they are intermediary to those shown.
In segments (a) and (b) the real and
imaginary parts of the central potentials are displayed
while
segments (c) and (d) respectively contain
the real and imaginary components
of the spin-orbit potentials.
The central potential values 
progressively becoming less refractive and more
absorptive with increasing energy.
The spin-orbit potentials have more variation
in their structure
with energy although there is a general
shape for the real and imaginary components.

Clearly while these components of the inversion
potentials vary
smoothly with energy, 
the central terms especially,
their shape is not characteristic of traditional
potentials, e.g. a Woods-Saxon function and/or its derivative.
That is so independent of the inherent pole
term in the inversion method 
which dominates the inversion potential
values  at the 
origin. But, as noted previously~\cite{MEpaper},
that pole influence is minimal beyond about 0.5 fm. Also
at each energy, the effect of the pole
when the inversion potential
was  used in calculation of scattering observables 
was small if not negligible.
Of primary interest then is the behavior of the inversion potentials 
from approximately 1 fm and our comments pertain to
the properties from that radius out.

In Fig.~\ref{pots},
 the real components of the central potentials found by inversion are shown
in panel (a).
At 65 MeV this component varies almost linearly between 
1 and 4 fm in radius before
tapering to zero, and then rather slowly. 
With increasing energy some structure develops in the central real
potential 
for radii $1 < r < 4$ fm
eventually, at 250 MeV, becoming
 a shoulder shaped well.
While values of the real
central potential in that region ($1 < r < 4$ fm)
decrease 
in strength to form the shoulder
shape, the longer range property increases in effect
with energy.

There is a different trend in the imaginary components of
 the central inversion potentials
as is evident in segment (b) of  Fig.~\ref{pots}. 
At all energies the long ranged character of the
central imaginary interactions
($r > 3.5$ fm) is the same and the component
rapidly vanishes with large $r$, 
contrasting markedly with
 the properties of
the associated central real terms. 
With increasing energy the central absorption also increases markedly
in the body of the nucleus ($1 < r < 3.5$ fm in this case) 
with a noticeable structure in the 65 MeV
potential result.
That absorptive character however is almost linear at 250 MeV.

The real and imaginary components of the spin-orbit
 potentials found by inversion are shown in
segments (c) and (d) of
Fig.~\ref{pots}. 
They are relatively weak,
the imaginary parts especially so
save for the result at 65 MeV. 
Overall these potentials are very similar; the real parts having a
weak attractive well (about 2.5 MeV deep at $r \sim 1.8$ fm) 
with a shorter ranged repulsion. 
No serious import should be attached as yet to the
specifics of the structures shown, 
other than that they
are the result of using the particular input set of 
phase shifts. Possible adjustments to the
basic $NN$ effective interactions in the original
non-local potential generation could  vary 
what we are to use as input
phase shifts sufficiently  to alter the small magnitude
details of the spin-orbit components shown here.

\subsection{The Phase Shifts}

\begin{figure}
\centering\epsfig{figure=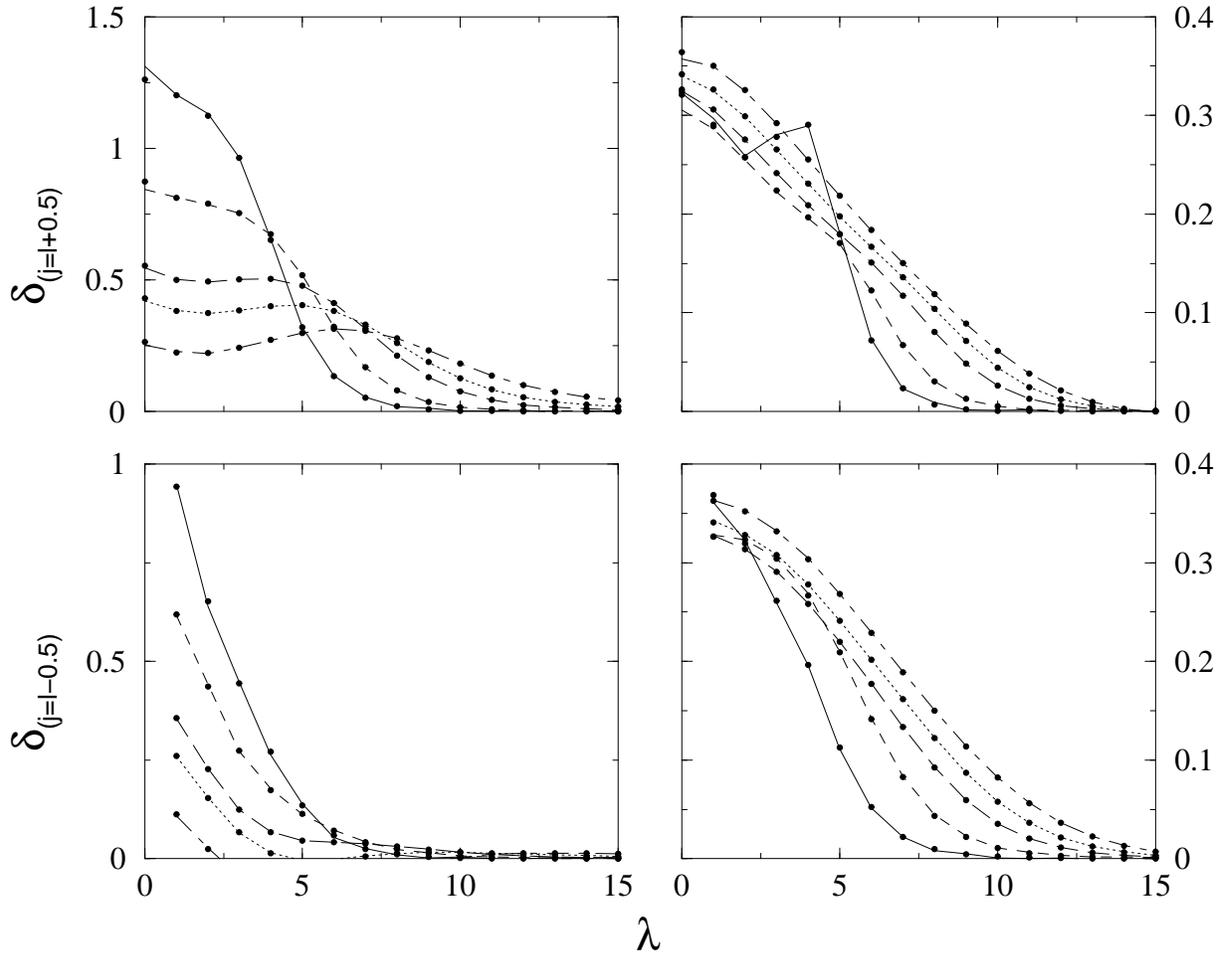,width=\linewidth,clip=}
\caption[]{
The phase shifts in radians
(real parts on the left, imaginary parts on the right)
obtained using the inversion potentials
given in Fig.~\ref{pots} compared with the
values used as input to the inversion procedure (filled dots).
The lines connecting
the physical values (integer $l$) identify
the energies with the same scheme as used in Fig.~\ref{pots}.  
}
\label{phases}
\end{figure}

Two potentials may be considered equivalent if they give
the same scattering
phase shifts. In principle 
the inversion potentials will be so.
Nevertheless, the inversion 
potentials 
must be used to get `inversion' phase shifts 
for comparison against the original values to check that they
are indeed equivalent.
In Fig.~\ref{phases} these  sets are compared 
for all energies. 
The real and imaginary parts of the phase shifts values
are shown in the left and right panels respectively
while those 
for $j = \ell + 0.5$
and
for $j = \ell - 0.5$
are given in the top and bottom sections respectively.
The input phase shift values are depicted by the filled circles 
while the results
found from the inversion potentials are indicated by
the connecting lines.
The notation with energy is that used in Fig.~\ref{pots}.
Clearly there is excellent agreement between the
phase shifts found from the
local, inversion potentials and from the $g$-folding non-local ones.

There is a relatively smooth and
consistent change
in the phase shift values as the energy increases.
The real parts of the sets corresponding to $j = \ell + 0.5$
and $j = \ell - 0.5$ both show a steady decrease in value with energy for
small partial waves, while the associated
imaginary parts show a steady increase. 
The notable exception is the 65 MeV case
for which the imaginary parts differ from this pattern. 
In the case of the imaginary
parts of the  $j = \ell + 0.5$ phase shifts
there is an unusual structure in the 65 MeV data set 
for the values $\ell < 5$.

A closer inspection of the results 
revealed that the s-wave phase
shifts
are slightly different from the original input values in both
the 65 MeV and the 100 MeV cases. 
This variation, however, is less than four percent.
The fact that
the phase shifts corresponding to $\ell=0$ differ is not unusual since
the centrifugal barrier screens all other partial wave
solutions from any (small) effect of the pole terms inherent 
in the inversion potentials.

\subsection{The cross sections and analyzing powers}

Although
the excellent agreement between phase shift sets obtained 
from the local (inversion)
and from original non-local
(full folding optical) potentials
for the p-$^{12}$C scattering at diverse energies is convincing,
another way to demonstrate this equivalence is to compare the 
associated observables.
By so doing any small variation in phase shifts that may exist 
can be emphasized. 
This is so as the cross
section spans several orders
of magnitude and then small inaccuracies within the phase shifts 
could be very apparent at the larger scattering
angles particularly.  The analyzing power likewise
should be  sensitive to small differences in the phase shift
values since that observable is given by differences between
scattering probabilities and is normalized by the differential
cross-section values. 

\begin{figure}
\centering\epsfig{figure=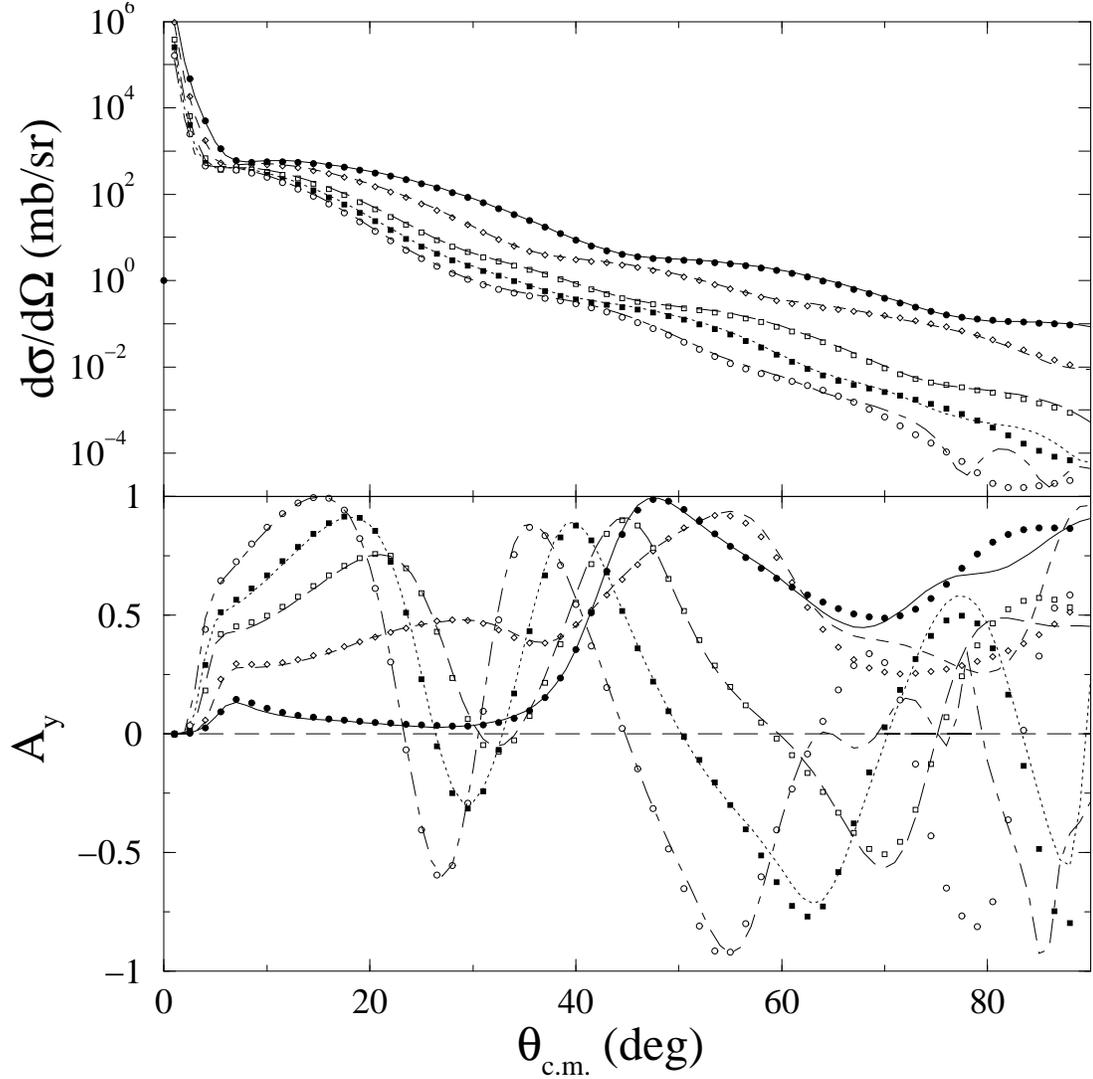,width=\linewidth,clip=}
\caption[]{
The differential cross sections (top) and analyzing powers (bottom)
obtained from use of the inversion potentials of
Fig.~\ref{pots} compared with the values 
associated with the phase shifts used as input to the inversion
procedure.
The lines indicate the disparate energy values as
specified in the text as do diverse symbols for the `data'. 
}
\label{xsecana}
\end{figure}

The cross sections and analyzing powers at the set of
energies chosen (65 to 250 MeV)
are given in Fig.~\ref{xsecana}.
In the top section of 
Fig.~\ref{xsecana}, 
the cross sections for each energy are displayed. 
Once again the values of the
the cross sections obtained from the full microscopic $NN$ folding 
non-local potentials
are represented by the circles and squares and the results 
obtained by
using the 
inversion potentials are portrayed by the diverse lines.
 The coding of those lines for each energy is as used in
Fig.~\ref{pots}.
Circles and squares, filled and open, have been used
to display the `data' to differentiate the 
set for each energy.
These cross section results demonstrate that the 
inversion potentials are
very good local equivalents. The cross section reproductions span eight
orders of magnitude and only for magnitudes less than 
$10^{-2}$mb/sr are small divergences evident.
Such divergence is most evident with
the 200 MeV and 250 MeV cases since those cross sections 
decrease most rapidly.

An even finer test of the agreement based upon observables between 
the local and non-local potentials are the
results for the analyzing powers. 
As shown in the bottom section of Fig.~\ref{xsecana}, 
reproduction of the analyzing power for each
energy is very good out to a 
center of mass scattering angle of 
$60^\circ$.
Even then only the cases of
65 and 100 MeV have any noticeable divergence between 
the local and non-local potentials results.
It is surmised that the small variations in the values 
of low-$\ell$ phase shifts are the cause
as such variations 
little effect predictions for the cross sections.
We conjecture
that this
behavior is a result of the choice 
we have made for
the phase shift value at the (unphysical) angular 
momentum,
$\ell = -0.5$; a quantity required in
the inversion process. 


\section{Frahn-Lemmer forms from local energy dependent potentials}

The problem is to use local forms of complex potential
$V_{\mathrm{LEQ}}({\bf r}; E) = V({\bf r};E)$ which when used in the
Schr\"odinger equation,
\begin{equation}
\frac{\hbar^2}{2\mu} \nabla^2 \phi({\bf k}, {\bf r}) 
+ \left[ E - V({\bf r}; E) \right] \phi({\bf k},{\bf r}) = 0 \; ,
\label{SEfulllocal}
\end{equation}
give phase shifts equal to those found by solution of the
non-local equations,
\begin{equation}
\frac{\hbar^2}{2\mu} \nabla^2 \chi^{(+)}({\bf k}, {\bf r}) 
+ E \chi^{(+)}({\bf k}, {\bf r}) - \int U({\bf r}, {\bf r'}; E)\:
\chi^{(+)}({\bf k}, {\bf r'})\: d{\bf r'} = 0 \; ,
\label{SEnonlocx}
\end{equation}
where that non-local form is of the Frahn-Lemmer type~\cite{Frahn}.

If the range of non-locality in $U({\bf r}, {\bf r'}; E)$
is small, then to
evaluate the integral term in the general form,
Eq.~\ref{SEnonlocx},
of the
Schr\"odinger equation, it is not necessary to know the
solution function $\chi^{(+)}(\bf{k},\bf{r})$ at all positions.
One only needs to know how $\chi^{(+)}(\bf{k},\bf{r})$ varies in a
volume element characterized by a small distance $`\bf{s}'$ about the
point $`\bf{r}'$.  In that volume element,
$\chi^{(+)}(\bf{k},\bf{r})$ oscillates with a wave number
$\bf{K}(\bf{r})$ so that integration over $`\bf{s}'$ will select
only those momentum components of any kernel that are in the
neighborhood of $\bf{K}(\bf{r})$.  This justifies expansion of
the Fourier transform of a kernel $G({\bf s})$ about the
local wave number,
\begin{equation}
G({\bf s}) = \frac{1}{(2\pi)^3} \int \tilde{G}(p) e^{-i{\bf p \cdot s}} 
d{\bf p}\ ,
\end{equation}
and retention of only the first two terms in the expansion,
\begin{equation}
\tilde{G}(p) = \tilde{G}(p^2) = \tilde{G}(K^2) + (p^2 - K^2)
\frac{d}{d (K^2)} \tilde{G}(K^2) + \cdots \; .
\end{equation}
In a simple manner then a local equivalent
potential to the exchange term in Eq.~(\ref{SEnonlocx}) can be
obtained by a Taylor series expansion,
\begin{eqnarray}
\int U( \bf{r}, \bf{r}'; E ) \chi^{(+)}(\bf{k}, \bf{r}') \:
d\bf{r}' & \approx \int U( \bf{r}, \bf{r}'; E ) e^{i(\bf{r'} -
\bf{r}) {\bf \cdot \nabla}} \chi^{(+)}(\bf{k},\bf{r} ) \: d\bf{r}'
\nonumber \\
& = \left[ \int U( \bf{r}, \bf{r}'; E ) e^{ i ( \bf{r}' - \bf{r} )
{\cdot \bf \kappa} } d\bf{r}'\right] \; \chi^{(+)}(\bf{k},\bf{r} )
\end{eqnarray}
where a local wave number,
\begin{equation}
\kappa(r) = \sqrt{2\mu \left[ E - V(r;E) \right]}\; ,
\end{equation}
has replaced the gradient operator.

Frahn and  Lemmer~\cite{Frahn}
assumed that the non-local kernels of the full Schr\"odinger
equation, Eq.~(\ref{SEnonlocx}), have a separable form
(this is also known in the literature as the 
Perey-Buck prescription),  
\begin{equation}
U({\bf r}, {\bf r'}; E )\: \to \: F({\bf R})\, 
v({\mbox{\boldmath $ \rho$}})\; ,
\label{FrahnLem}
\end{equation}
where
\begin{equation}
{\bf R} = \frac{1}{2} ( {\bf r} + {\bf r'} ) 
\ \ ;\ \  
{\mbox{\boldmath $\rho$}} =
{\bf r} - {\bf r'} \; .
\end{equation}
Furthermore, they
assume that $F({\bf R}) = F(R)$ with $F(R)$ a slowly varying function
about $R = r$, and that
\begin{equation}
v({\mbox{\boldmath$ \rho$}}) 
= v(\rho) = \left(\pi \sigma^2\right)^{-\frac{3}{2}}
\exp{\left( -\frac{\rho^2}{\sigma^2} \right)}\ ,
\label{FLnon-loc}
\end{equation}
where $\sigma$ is the non-locality range.  A Taylor expansion about
$`\bf{r}'$ gives to second order
\begin{eqnarray}
&&F(R) \chi^{(+)}( \bf{k}, \bf{r}' ) \approx 
\nonumber\\
&&\hspace*{0.5cm}
F(r) \chi^{(+)}( {\bf k}, {\bf r} ) + \frac{1}{6} \rho^2 \left\{
\frac{d}{dr}F(r)\frac{d}{dr} + F(r) \nabla^2 + \frac{1}{4r}
\frac{d^2}{dr^2} [ rF(r) ] \right\} \chi^{(+)}( {\bf k}, {\bf r} )
\label{Tayexp}
\end{eqnarray}
and this gives
\begin{eqnarray}
&&\int U( {\bf r}, {\bf r'}; E ) \chi^{(+)}( {\bf k}, {\bf r'} ) \:
d{\bf r'} \approx 
\nonumber\\
&&\hspace*{0.5cm}
\left\{ v_0 F(r) + \frac{1}{8} v_2 \frac{1}{r} \frac{d^2}{dr^2} \left[
rF(r) \right] + \frac{1}{2} v_2 \frac{d}{dr}F(r) \frac{d}{dr} +
\frac{1}{2} v_2 F(r) \right\} \chi^{(+)}( \bf{k}, \bf{r} ) \; ,
\end{eqnarray}
where $v_n$ are the moments of the
non-locality,
\begin{equation}
v_n = \frac{4\pi}{(2n -1)} \int_0^\infty v(\rho) \rho^{(2+n)} \: d\rho
\; .
\end{equation}
Under this approximation, the non-local Schr\"odinger equation
reduces to
\begin{eqnarray}
&&\left\{ \left[ -\frac{\hbar^2}{2\mu} + \frac{1}{4}\sigma^2 F(r)
\right] \nabla^2 + \frac{1}{4} \sigma^2 \frac{d}{dr} F(r) \frac{d}{dr}
\right. 
\nonumber\\
&&\hspace*{1.5cm}
\left. \mbox{} - E + F(r) + \sigma^2 \frac{1}{r} \frac{d^2}{dr^2}
\left[ rF(r) \right] \right\} \chi^{(+)}( {\bf k}, {\bf r} ) = 0 \; ,
\label{SEnonloc3}
\end{eqnarray}
which maps to an equivalent local and energy dependent form,
\begin{equation}
\left[ -\frac{\hbar^2}{2\mu} \nabla^2 + V(r, E) - E \right]
\varphi({\bf k}, {\bf r}) = 0 \; ,
\label{SEFLloc}
\end{equation}
upon using a point transformation defined by
\begin{equation}
\varphi({\bf k}, {\bf r}) = T(r) \chi^{(+)}({\bf k}, {\bf r}) \; .
\label{transform}
\end{equation}
Multiplication of Eq.~(\ref{SEnonloc3}) by $T(r)$ gives
\begin{equation}
T(r) \left\{ -\frac{\hbar^2}{2\mu}\nabla^2 - \left[
\frac{J_1(r)}{X(r)} \right] \frac{d}{dr} \left[ \frac{J_0(r) +
E}{X(r)} \right] \right\} T^{-1}(r) \varphi({\bf k}, {\bf r}) = 0
\; ,
\label{Tmod}
\end{equation}
with the functions,
\begin{eqnarray}
J_0(r) &=& F(r) + \frac{\sigma^2}{16r} \frac{d^2}{d r^2} \left[ rF(r)
\right]\; ,
\nonumber\\
J_1(r) &=& \frac{\sigma^2}{4} \frac{d}{d r} F(r) \; ,
\nonumber\\
X(r) &=& 1 - \frac{\mu\sigma^2}{2\hbar^2}  F(r) \; .
\label{JXeqn}
\end{eqnarray}
The first derivative term is eliminated by choosing $T(r) =
\sqrt{X(r)}$ as then Eq.~(\ref{Tmod}) becomes
\begin{equation}
\left[ -\frac{ \hbar^2 }{ 2\mu }\nabla^2 
+ \frac{\Upsilon(r) - E }{X(r)} \right] 
\varphi({\bf k}, {\bf r}) = 0 \; ,
\end{equation}
so identifying
\begin{equation}
V(r, E) = E + \frac{\Upsilon (r) - E}{X(r)}
\label{Uleq}
\end{equation}
when
\begin{eqnarray}
\Upsilon(r) &=& J_0(r) - \frac{1}{r}J_1(r) - \frac{1}{2}
\frac{d}{dr}J_1(r) - \frac{3}{2} \frac{1}{X(r)} J^2_1(r) \nonumber \\
&=& F(r) - \frac{\sigma^2}{8r} \frac{d}{dr} F(r) -
\frac{\sigma^2}{16r} \frac{d^2}{dr^2} F(r) -
\frac{3\mu\sigma^4}{32\hbar^2} \frac{1}{ \left[ 1 - \mu\sigma^2/\left(
2\hbar^2 \right) \right] F(r) } \left[ \frac{d}{dr} F(r) \right]^2
\label{align}
\end{eqnarray}

Following Apagyi {\it et al.}~\cite{Apagyi}, by considering local
equivalent potentials at two energies
$E_1$ and $E_2$, it follows by using Eq.~(\ref{Uleq}) that
\begin{equation}
\Upsilon(r)
=
\frac{E_2 V(r,E_1) - E_2 V(r,E_2)}
{\left[ V(r,E_1) - V(r,E_2) + E_2 - E_1 \right]}\ ,
\label{Upkeq}
\end{equation}
and that
\begin{equation}
F(r) = \frac{\hbar^2}{\mu}
\frac{2}{\sigma^2}
\frac{V(r,E_1) - V(r,E_2)}
{\left[ V(r,E_1) - V(r,E_2) + E_2 - E_1 \right]}\ .
\label{Feqn}
\end{equation}
Then from Eqs.~(\ref{JXeqn}),
one can find $J_0(r), J_1(r),$ and $X(r)$,
and so have a complete specification of the non-local
interaction properties.
To the extent that the energy dependent Frahn-Lemmer form
describes p-${}^{12}$C scattering the functions,
$U(r)$, $F(r)$ and $\Upsilon(r)$ should not be energy dependent.
As will be seen, that is not completely the case
with the system we have studied.


\section{The Frahn-Lemmer potentials for $p-{}^{12}$C}

The central parts of the inversion potentials
that were specified starting with 
phase shift sets from the $g$ folding optical 
potential calculations of p-${}^{12}$C
scattering at 100, 135, 160, and 200 MeV
have been used in this study.
With the non-locality range, $\sigma$,
 taken first as 0.7 and subsequently  as 1.0 fm,
 the pairs of inversion potentials
with energies 100 and 135 MeV, with 135 and 160 MeV,
 and with 160 and 200 MeV, have been used 
 to 
find the functions $U(r)$, $F(r)$, and $\Upsilon(r)$ that characterize
the non-local Frahn-Lemmer form of the p-${}^{12}$C optical potential.
Results found using those pairs are identified 
by the notation 100-135, 135-160, and 160-200,
and are portrayed in the next three figures by the solid,
long dash, and short dash curves respectively.
The real and imaginary parts of the various functions are given
in the top and bottom sections of these figures with the results
found using $\sigma$ = 0.7 and 1.0 fm presented
in the left and right side panels respectively.

\begin{figure}
\centering\epsfig{figure=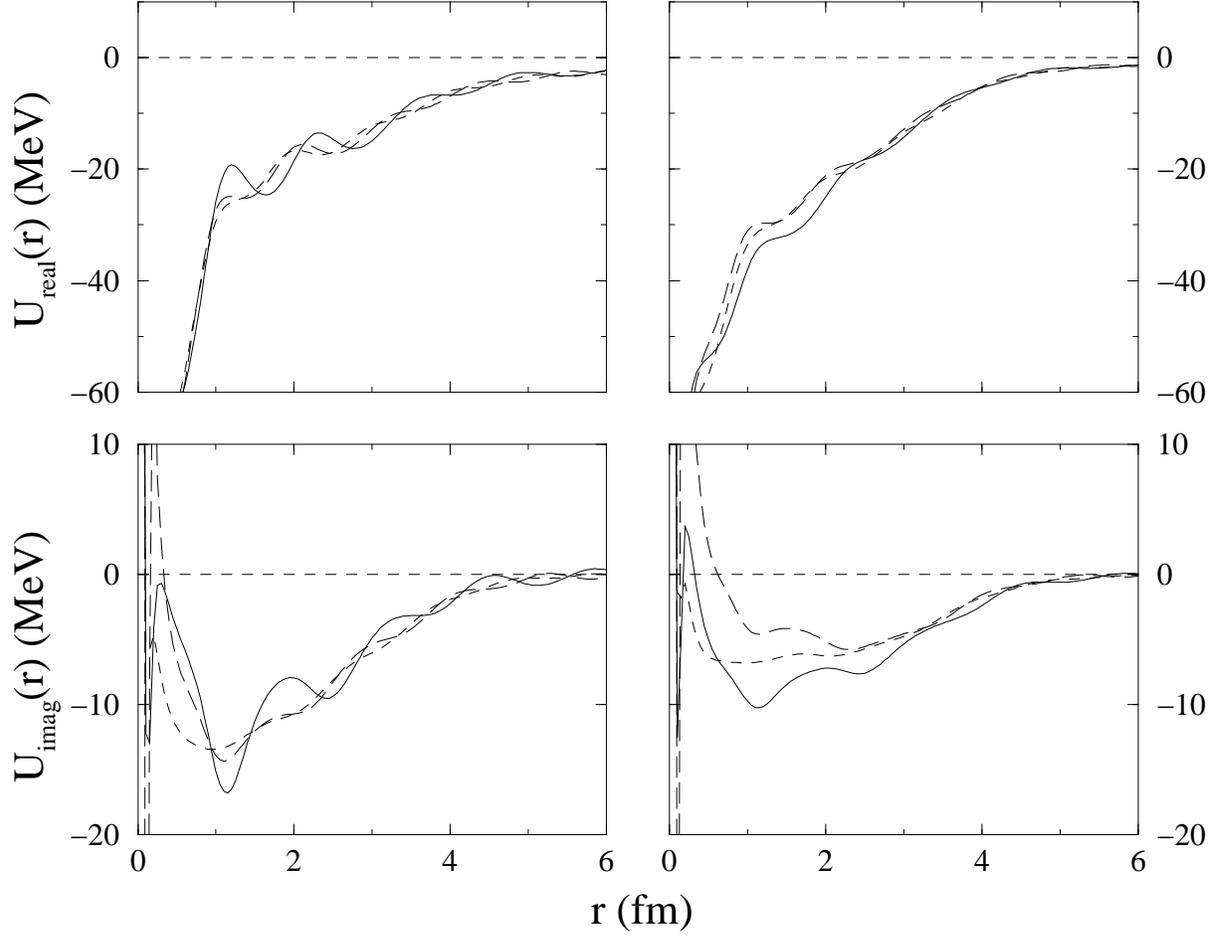,width=\linewidth,clip=}
\caption[]{The local components, $U(r)$ of Eq.~(\ref{SEnonlocx}),
obtained with non-locality ranges, $\sigma$, of 0.7 and 1.0 fm
(left and right panels respectively) and as deduced from the 
100-135 MeV (solid curves),
the 135-160 MeV (long dash curves), and the 160-200 MeV
(short dash curves) pairs of (local) inversion potentials.
}
\label{UwithE}
\end{figure}

\begin{figure}
\centering\epsfig{figure=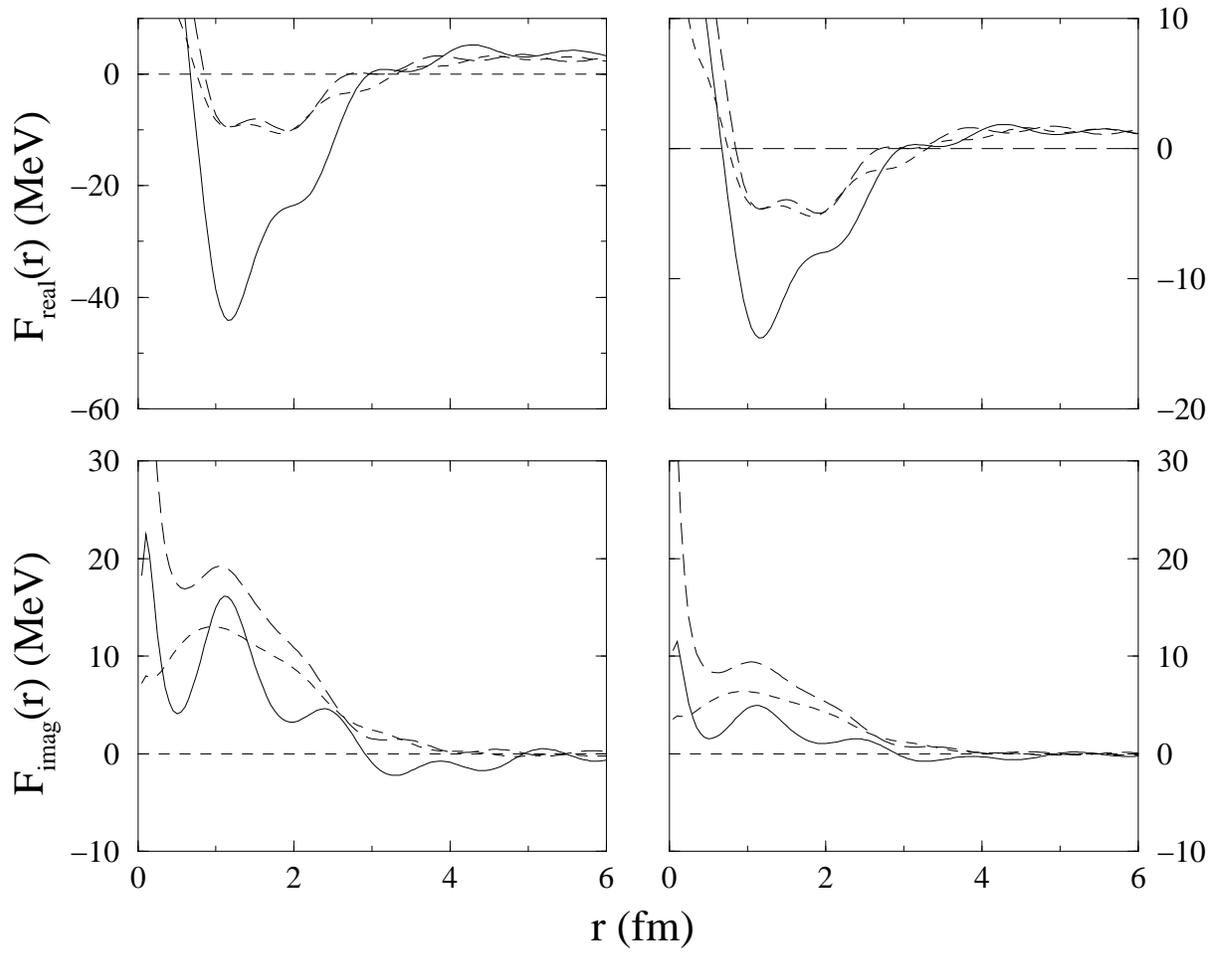,width=\linewidth,clip=}
\caption[]{
As for Fig.~\ref{UwithE}, but for the non-local components, $F(r)$.
}
\label{FwithE}
\end{figure}

\begin{figure}
\centering\epsfig{figure=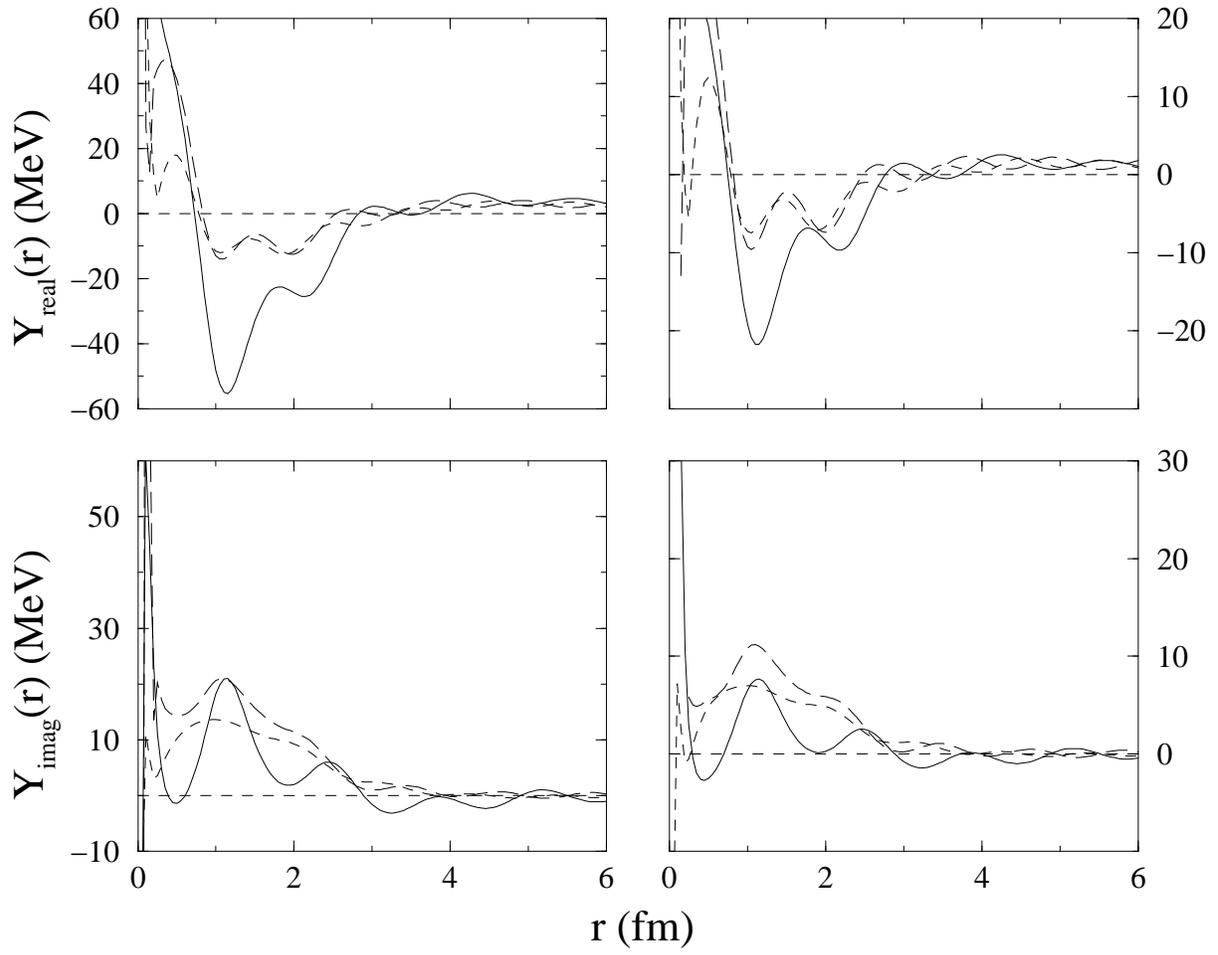,width=\linewidth,clip=}
\caption[]{
As for Fig.\ref{UwithE}, but for the 
non-local components, $\Upsilon(r)$. 
}
\label{KwithE}
\end{figure}

The results for the local attribute $U(r)$ 
of the Frahn-Lemmer representations for the non-local 
optical potential
are given in Fig.~\ref{UwithE}.
In general those components are similar for all three energy pairs,
and more so for the 135-160 and 160-200 cases. 
The variations in these results are but a few MeV in size.
However there is a noticeable change in the degree of structure
with marked oscillations in the results found with the smaller
(0.7) non-locality range.
Still the results with both non-locality range values do exhibit
a residual
energy dependence, and such is presumed not to be the case with the 
Frahn-Lemmer
prescription.
The modulating functions, $F(r)$, of the actual non-local term
in the Frahn-Lemmer form 
are displayed in Fig~\ref{FwithE}.
The real parts of this function are quite similar in the 135-160 and
 160-200 cases although the overall strength of the real part of $F(r)$
decreases with the energy as it does
with increase in the non-locality range.
The 100-135 results
(for the real part of $F(r)$)
 are larger than the others 
and varies from those in
structure.
The imaginary parts of $F(r)$ vary noticeably
with the marked structure of the 100-135 results
diminishing with energy.  But the size change is not linear.
Such energy dependence is also at odds with the Frahn-Lemmer 
requirement of an energy independent non-local potential.
The local elements $\Upsilon(r)$ 
for these cases are shown in Fig.~\ref{KwithE}.
They strongly reflect the properties of the relevant $F(r)$
albeit that the structures are enhanced.

\begin{figure}
\centering\epsfig{figure=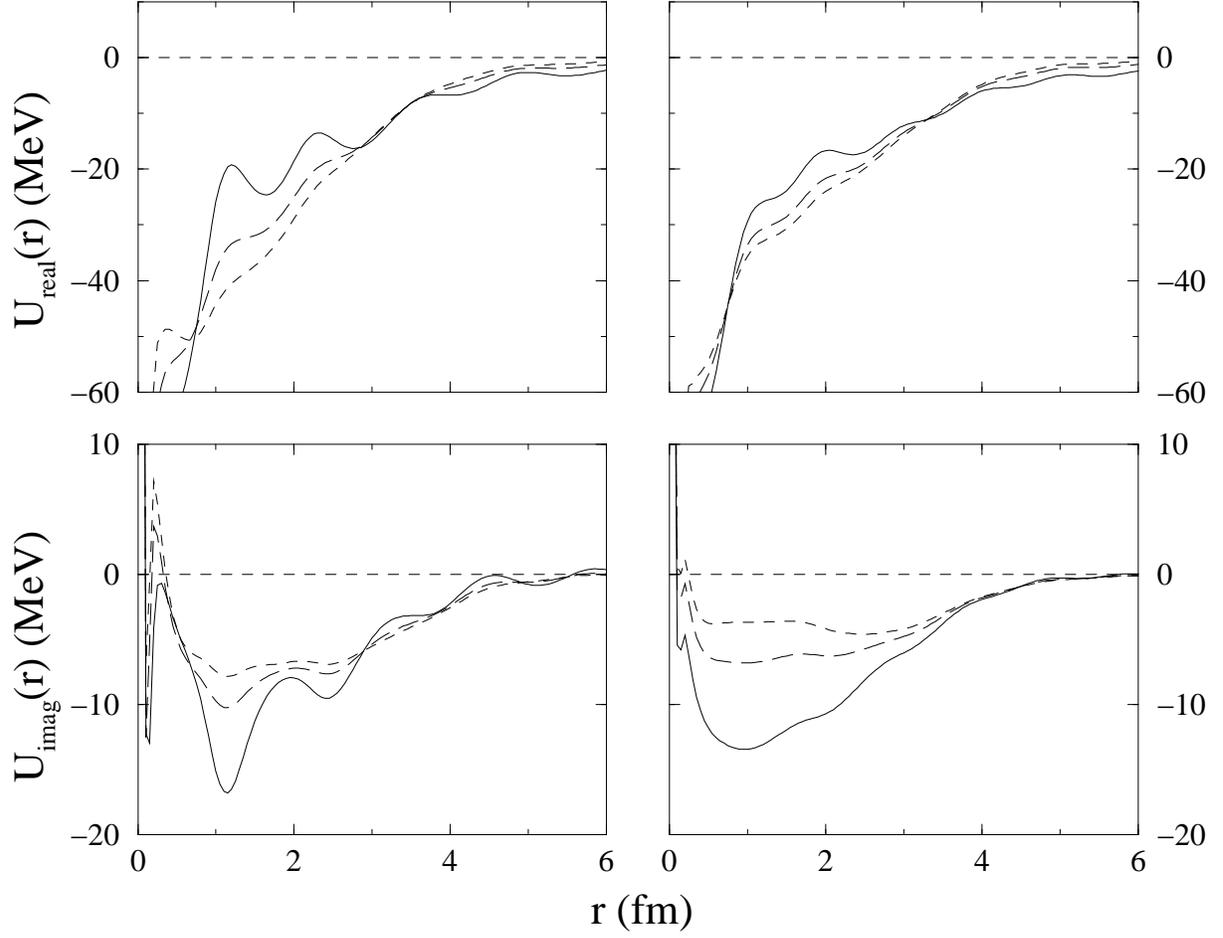,width=\linewidth,clip=}
\caption[]{The local terms $U(r)$ 
found using the 
100-135 MeV (left)
and the 160-200 MeV (right)
pairs of inversion potentials but for 
Frahn-Lemmer non-locality ranges
of 0.7, 1.0, and 1.4 fm. The results
are portrayed by the solid, long dash, 
and short dash curves respectively.
}
\label{UwithS}
\end{figure}

\begin{figure}
\centering\epsfig{figure=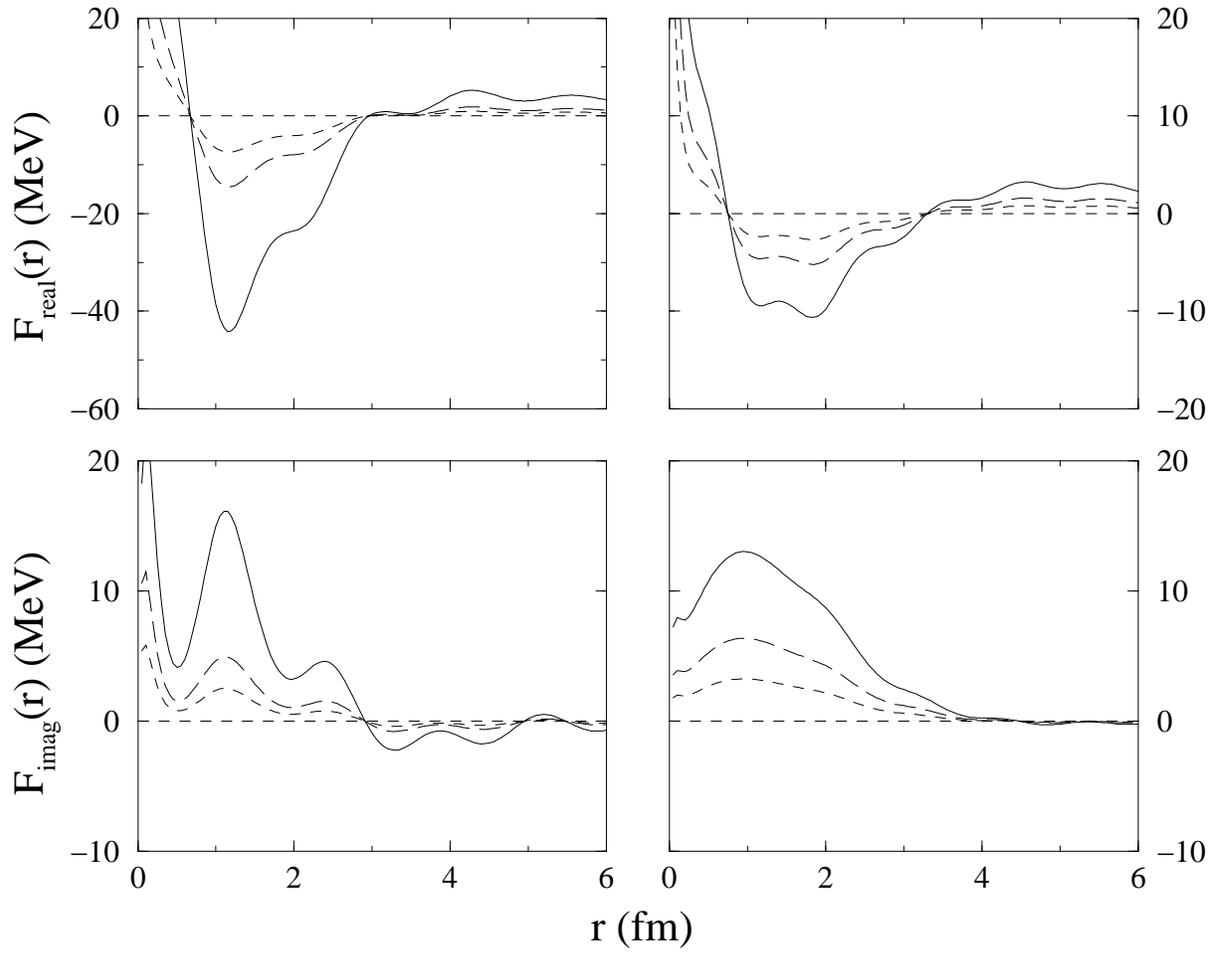,width=\linewidth,clip=}
\caption[]{
As for Fig.~\ref{UwithS}, but for the non-local terms, $F(r)$.
}
\label{FwithS}
\end{figure}

\begin{figure}
\centering\epsfig{figure=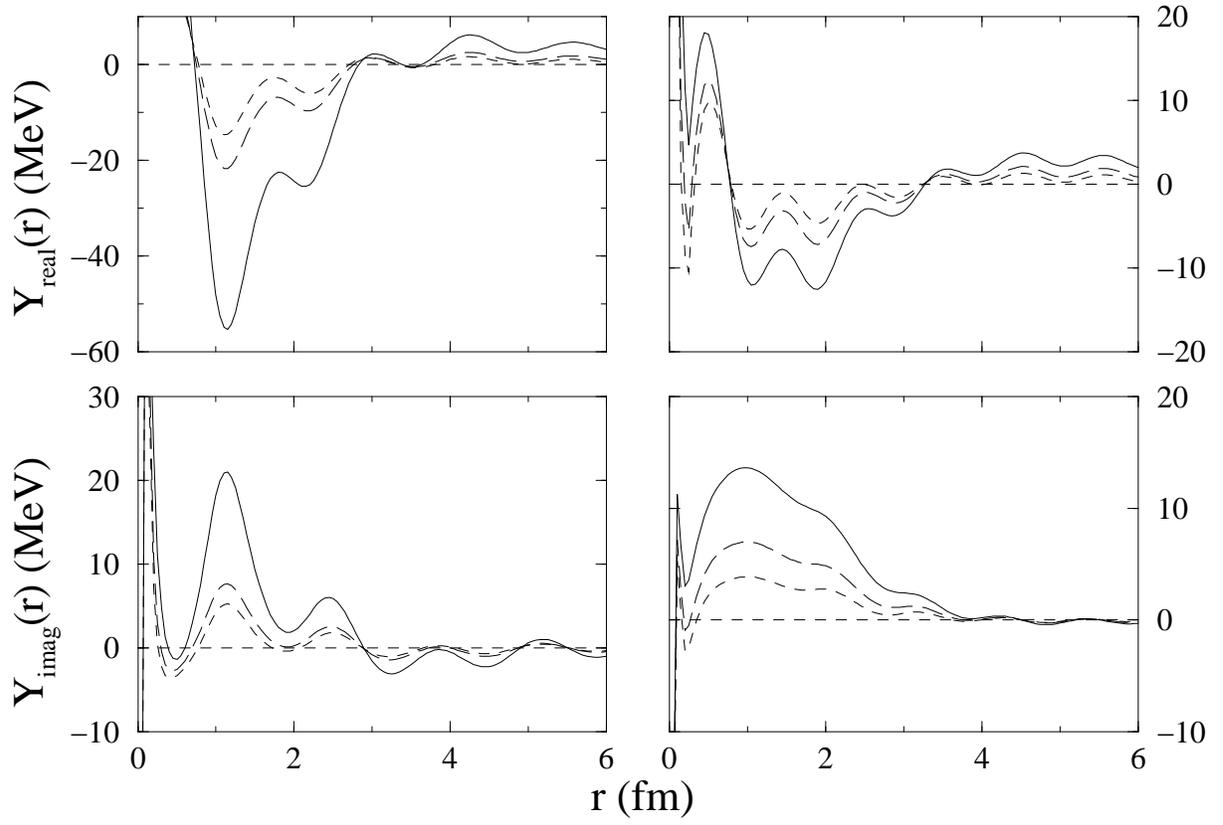,width=\linewidth,clip=}
\caption[]{
As for Fig.~\ref{UwithS}, but for the non-local terms, $\Upsilon(r)$.
}
\label{KwithS}
\end{figure}

In the next three figures the
100-135 and the 160-200 inversion potential
pairs are shown again but now
to compare more directly the effects
of different choices of 
 the non-locality range, $\sigma$.
Results found with $\sigma$ =
0.7, 1.0, and 1.4 fm now are displayed by the solid,
long dash, and short dash curves respectively. 
Again the real and imaginary parts
of the characteristic functions $U(r)$, $F(r)$, and $\Upsilon(r)$ are
given in the top and bottom sectors of the diagrams.
The $U(r)$ components are displayed in Fig.~\ref{UwithS}.
The real parts for both energy pairs are similar with 
a decrease in the structure
of the results being evident as the non-locality range increases.
 Indeed the results with $\sigma$ = 1.0 and 1.4 fm are very similar.
The imaginary parts of $U(r)$ have the form of an attractive well.
Again the structure observed with the shortest range washes out
 when $\sigma$ is increased.  Now also the 160-200 MeV
result changes strength noticeably with increase of the non-locality range.
The results for $F(r)$ and $\Upsilon(r)$ are
given in Figs.~\ref{FwithS} and \ref{KwithS}
respectively.  The $\Upsilon(r)$ variations reflect those of $F(r)$ as before.
In these cases changing the non-locality range has a dramatic effect, 
largely upon  the magnitudes.
In part though, that
might  be considered just an off-set
to the normalization which depends on $\sigma$.
But there are also changes in the structures.
The disparity between the 100-135 and 160-200
results emphasizes again that there is a residual
energy dependence one must consider if the 
set of fixed energy inversion potentials
are recast as a Frahn-Lemmer type of non-local interaction.


\section{Conclusions}

The fixed energy inverse scattering 
method of Lun {\it et al.}~\cite{Lun}
has been used to specify local potentials
from sets of phase shifts given by solutions
of the Schr\"odinger equations with 
non-local optical potentials for proton-${}^{12}$C
elastic scattering for a range of proton energies.
Those (non-local potential)
phase shift sets give very good fits
to observed cross-section and analyzing power data~\cite{Deb}.
So also then do the local (inversion) potentials 
as we have shown them
to be both phase shift and observable equivalent
(to the non-local potential expectations) to
a high degree. 
As a non-local to local potential conversion scheme,
the inverse scattering theory method has
proved to be very effective. 
Irrespective of the pole term inherent in the inverse scattering theory
method, the local inversion potentials do not resemble
the simple functional forms, e.g. Woods-Saxon potentials,
that are commonly used in phenomenological
(numerical inversion) analyses of such
scattering data.

We then considered a reverse mapping to see
if and how  the energy dependence of the inversion potentials
might reflect a non-locality of simpler functional
kind, and of the Frahn-Lemmer form in particular.
The results are indicative of 
a characteristic forms for the diverse
components of that
simple non-local form of the interaction,
but the detailed properties can vary significantly with 
the choice of the non-locality range,
and there is an energy dependence
residual in the Frahn-Lemmer functions.
Thus we contend that the energy dependence 
of local potentials for p-${}^{12}$C scattering
(and by implication for other targets)
whether those potentials are found by inversion or
by phenomenology, is not solely a reflection
of the true non-locality in the interaction
between the nuclei.  The non-locality itself is also energy
dependent.



\end{document}